\immediate\write18{makeindex \jobname.nlo -s nomencl.ist -o \jobname.nls}
\pdfoutput=1
\documentclass[preprint, 5p]{elsarticle}
\usepackage{amsmath}
\usepackage{booktabs}
\usepackage{array}
\usepackage[capitalise,nameinlink]{cleveref}
\usepackage{siunitx}
\usepackage{graphicx}
\usepackage{framed}
\usepackage{multicol}
\usepackage{multirow}
\usepackage{nomencl}
\usepackage{color}
\usepackage{afterpage}
\usepackage[labelfont=bf,justification=raggedright,singlelinecheck=true]{caption}
\usepackage[defaultlines=4,all]{nowidow}

\graphicspath{{./images/}}
\makenomenclature
\renewcommand*\nompreamble{\begin{multicols}{2}}
\renewcommand*\nompostamble{\end{multicols}}

\biboptions{square,comma,numbers,sort}

\begin{document}

\journal{arXiV cs.CE}

\begin{frontmatter}

\title{A Semi-Analytical Method for Calculating Revisit Time for Satellite Constellations with Discontinuous Coverage}

\author[UoM]{Nicholas~H.~Crisp\corref{cor1}}
\ead{nicholas.crisp@manchester.ac.uk}

\author[UoM]{Sabrina~Livadiotti}
\ead{sabrina.livadiotti@postgrad.manchester.ac.uk}

\author[UoM]{Peter~C.E.~Roberts}
\ead{peter.c.e.roberts@manchester.ac.uk}

\cortext[cor1]{Corresponding author.}
\address[UoM]{School of Mechanical, Aerospace and Civil Engineering, The University of Manchester, George Begg Building, Sackville St, Manchester, M13~9PL}

\begin{abstract}
This paper presents a unique approach to the problem of calculating revisit time metrics for different satellite orbits, sensor geometries, and constellation configurations with application to early lifecycle design and optimisation processes for Earth observation missions. The developed semi-analytical approach uses an elliptical projected footprint geometry to provide an accuracy similar to that of industry standard numerical orbit simulation software but with an efficiency of published analytical methods. Using the developed method, extensive plots of maximum revisit time are presented for varying altitude, inclination, target latitudes, sensor capabilities, and constellation configuration, providing valuable reference for Earth observation system design.
\end{abstract}


\end{frontmatter}

\begin{table*}[!t]   
\begin{framed}
	\nomenclature{$R_a$}{Earth equatorial radius}
	\nomenclature{$R_b$}{Earth polar radius}
	\nomenclature{$R$}{Earth radius}
	\nomenclature{$\omega_E$}{Earth rotation rate}
	\nomenclature{$\phi$}{Target latitude}
	\nomenclature{$\lambda$}{Longitude}
	\nomenclature{$i$}{Inclination}
	\nomenclature{$\omega$}{Argument of perigee}
	\nomenclature{$h$}{Altitude}
	\nomenclature{$a$}{Semi-major axis}
	\nomenclature{$e$}{Eccentricity}
	\nomenclature{$p$}{Semilatus Rectum}
	\nomenclature{$\Omega$}{Right Ascension of Ascending Node (RAAN)}
	\nomenclature{$\nu$}{True anomaly}
	\nomenclature{$r_s$}{Orbital radius}
	\nomenclature{$\psi$}{Sensor boresight half cone angle}
	\nomenclature{$\rho$}{Slant range}
	\nomenclature{$\gamma$}{Intermediate angle}
	\nomenclature{$\epsilon$}{Elevation angle}
	\nomenclature{$\theta$}{Half ground range angle of sensor}
	\nomenclature{$\Lambda$}{Half surface dihedral angle of sensor}
	\nomenclature{$P_k$}{Keplerian period}
	\nomenclature{$P_n$}{Nodal period}
	\nomenclature{$\mu_E$}{Earth gravitational parameter}
	\nomenclature{$J2$}{Second-degree Earth zonal harmonic}
	\nomenclature{$n$}{Mean motion}
	\nomenclature{$\delta\Omega$}{Regression rate of RAAN}
	\nomenclature{$\Delta\lambda$}{Ground track shift}
\printnomenclature
\end{framed}
\end{table*}

\section{Introduction} \label{sec:Introduction}
During the design of Low Earth Orbit (LEO) systems of satellites performance metrics are required to evaluate different orbital configurations and payload specifications. For Earth observation or communications missions the extent of coverage of the system or rate at which the system visits or views different locations is of particular interest to the mission designer. Revisit time (also known as the response time or coverage gap) is often used as a key performance metric for LEO systems which do not have continuous coverage of an area of interest and is defined as the duration in time between consecutive viewings of a given location on the Earth. Most commonly, the Maximum Revisit Time (MRT) and Average Revisit Time (ART) over a given target area and period of analysis are considered during the mission design process.

With increasing on-orbit capability and reduced system cost, development of constellations of small satellites has recently grown significantly. However, the design and optimisation of these systems is complex and multidisciplinary, owing to the number of different design variables and ranges which they can take. For example, consideration of the orbit design, system configuration, payload characteristics, and mission performance can all have a significant effect on the final system utility and cost. For an Earth observation satellite or constellation the evaluation of revisit time therefore forms a critical component of the design and optimisation process.

Often the analysis of coverage or revisit metrics for satellites and constellations is performed using commercially available orbital propagation and simulation software such as STK (Systems Tool Kit)~\cite{Nunes2013,Marinan2013,Nag2015}. However, due to the numerical nature of these programs and potential for long analysis periods (on the order of minutes) or large numbers of satellites, the computational time can become considerable. Furthermore, when many cases are to be considered, for example within a wider framework for system optimisation, a faster, open, and stand-alone method is preferred.

A number of methods for calculating revisit time have been discussed and applied in the literature. \citet{Wertz1999,Wertz2011} provides basic descriptions for evaluating coverage using simple analytical expressions and calculating revisit metrics using two numerical treatments. The first, a numerical method, utilises simple ground track plots and mission geometry and is shown to be most useful for rapid mission analysis. The second, \textit{point coverage simulation} methods, use a grid of points at which visibility characteristics are be evaluated. These methods are able to provide insight into the statistical measures of coverage and can achieve greater accuracy albeit at the cost of longer computation times.

\citet{Bottkol1991} describe a numerical phase-based approach to the calculation of revisit interval. In this method the satellite ground track is mapped to the surface of a torus which is then unwrapped to indicate the intersection with a defined visibility region. The developed method is subsequently used to perform the design of satellite constellations for optimal coverage and revisit characteristics. However, the analysis is limited to orbits which have a short repeating ground track period and can therefore not address the complete range of low Earth orbits.

\citet{Ulybyshev2014,Ulybyshev2015} presents a geometric analysis method for the calculation of revisit time. Reasonable agreement on the order of 2\% error in maximum revisit time is shown with STK for a limited number of cases. \citet{Razoumny2016} also presents a general analytical method for the problem of discontinuous coverage. However, for these methods only limited validation is provided and computational performance is not given explicitly.

The design of satellite constellations for optimal revisit metrics has been studied extensively in the literature. \citet{Lang1983} first worked towards minimization of MRT by considering variable orbital inclination and constellation configuration. \citet{Lang2002} later applied a genetic algorithm optimisation process to this problem enabling a parametric exploration of the design space. \citet{Crossley2000} similarly investigated the use of genetic algorithm and simulated annealing approaches to minimisation of MRT of satellite constellations. Further investigations of satellite constellation optimisation for revisit and coverage metrics are performed by \citet{Ferringer2006,Ferringer2007}. However, in each of these studies the analysis of revisit time is performed using the proprietary COVERIT and ASTROLIB/Revisit-C programs of The Aerospace Corporation. These tools utilise a numerical \textit{point coverage simulation} method and generated ephemeris tables for the satellites to calculate revisit time. A representative computation time of \SI{0.5}{\second} per function call is given by \citet{Ferringer2006}  on an Intel Pentium III \SI{1200}{\mega\hertz} processor.

\citet{Abdelkhalik2006} address the design of optimal satellite orbits for the surveillance of multiple target sites. The method utilises propagated satellite orbits over a short time period and a penalty function method to design satellite orbits for maximum resolution or maximum observation time. \citet{Abdelkhalik2011} subsequently perform the design of optimal repeat ground track sun-synchronous orbits for multiple site surveillance by considering the intersection of the J2-perturbed rotating orbital plane with the target sites.

Given the noted interest in extensive trade-studies for multi-satellite systems and constellations, using large-scale design optimisation or Monte Carlo processes, a means to efficiently calculate orbital revisit metrics is necessary. Furthermore, the solution of observation characteristics for numerous orbital conditions can be beneficial for VLEO and CubeSat or nano-class missions which may consist of many payloads operating at low orbital altitudes and subject to free or controlled decay.

The method presented herein offers an accurate and efficient analytical calculation process for revisit metrics of individual satellites or constellations with discontinuous coverage of a given target observation area. Consideration of the oblate spheroid approximation for the Earth and associated orbital perturbation is included. The method is demonstrated to be capable of rapidly calculating orbital revisit time with an accuracy on the order of commercial numerical simulation software. Finally, revisit characteristics for different target latitudes, orbital parameters, sensor characteristics, and constellation configurations are explored.


\section{Calculation of Maximum Revisit Time}

Calculation of the revisit time at a given latitude can be determined by correlating the longitude of all projected passes and the instantaneous coverage of the sensor over a period of analysis. The maximum time gap between any two continuous passes (ordered by time) at any longitude gives the simple MRT. Similarly, ART can be calculated. Contrastingly, the time to 100\% coverage can be determined when the maximum difference in longitude between any two contiguous passes (ordered by longitude) falls below the angular range of the sensor.

To maintain the accuracy of this method for low inclination orbits and at increasing latitudes of interest consideration must be given to rotation of the Earth during each pass and the angle of the orbit track with respect to the latitude of interest. These effects can provide sensor access to longitudes which are beyond the simple sensor width at the point where the orbit intercepts the latitude of interest. These passes can be examined by considering the ground track of the orbit in the vicinity of the latitude crossing and by idealising the sensor coverage area as an ellipse.

\subsection{Simple Orbit and Sensor Geometry}

\begin{figure*}[bt]
	\centering
	\def\svgwidth{150mm}
	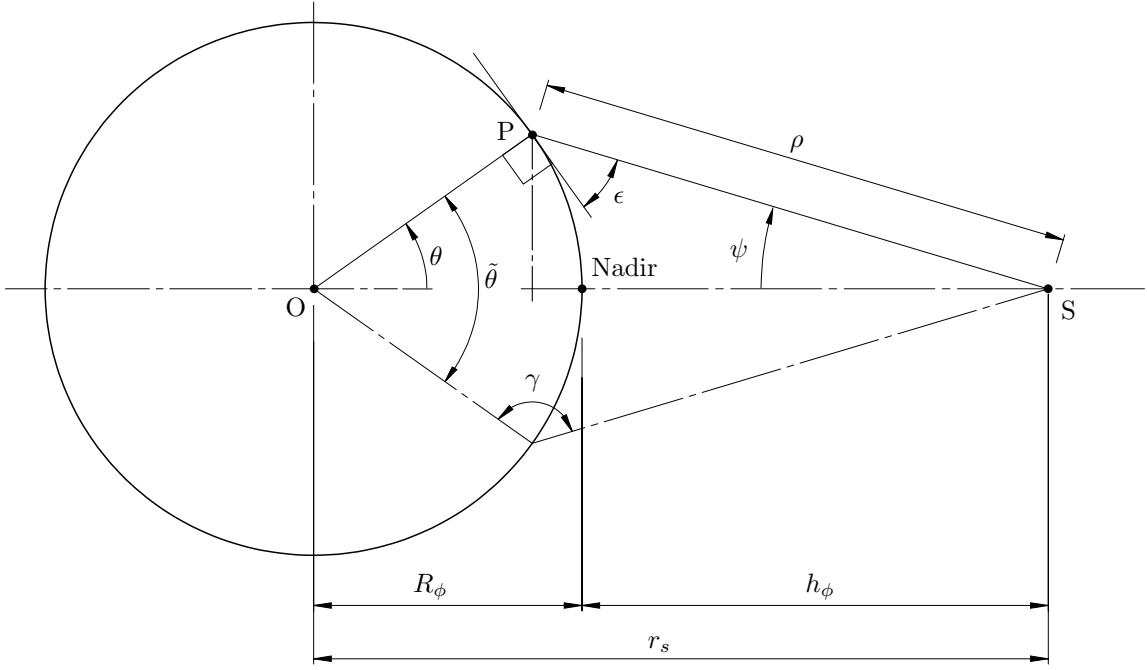
	\caption{Geometry of satellite sensor field of regard. Adapted from \citet{Vallado2013}.}
	\label{F:FoRGeometry}
\end{figure*}

The calculation of maximum revisit time first requires a treatment of the orbit geometry and Field of Regard of the sensor, shown in \cref{F:FoRGeometry}, to yield the half ground range angle $\theta$ visible to the satellite. Alternatively, a minimum elevation angle $\epsilon$ constraint can be applied to specify the sensor geometry and half ground range angle $\theta$.

First, the geodetic radius $R_{\phi}$ is calculated at the target latitude $\phi$. Here, the Earth is assumed to be an oblate spheroid where $R_a$ and $R_b$ are the equatorial and polar radii respectively.
\begin{equation} \label{E:Rad_gd}
R\left(\phi\right) = \sqrt{ \frac{\left(R_a^2\cos{\phi}\right)^2 + \left(R_b^2\sin{\phi}\right)^2}{\left(R_a\cos{\phi}\right)^2 + \left(R_b\sin{\phi}\right)^2} }
\end{equation}
The height $h_{\phi}$ and radius of the satellite $r_s$ at nadir can be found by considering the semi-major axis $a$, eccentricity $e$, argument of perigee $\omega$, and true anomaly $\nu_\phi$ at the target latitude.
\begin{equation} \label{E:TrueAnom}
\nu_{\phi} = \sin^{-1} \left\{\dfrac{\sin{\phi}}{\sin{i}} \right\} - \omega
\end{equation}
\begin{equation} \label{E:EllRadius}
r_s = R_\phi + h_{\phi} = \frac{p}{1 + e\cos{\nu_{\phi}}}
\end{equation}
where the semilatus rectum $p$ is defined as
\begin{equation} p = a(1-e^2) \nonumber \end{equation}

\begin{figure}[bt]
	\centering
	\begin{minipage}{\linewidth}
	\centering
	\def\svgwidth{110mm}
	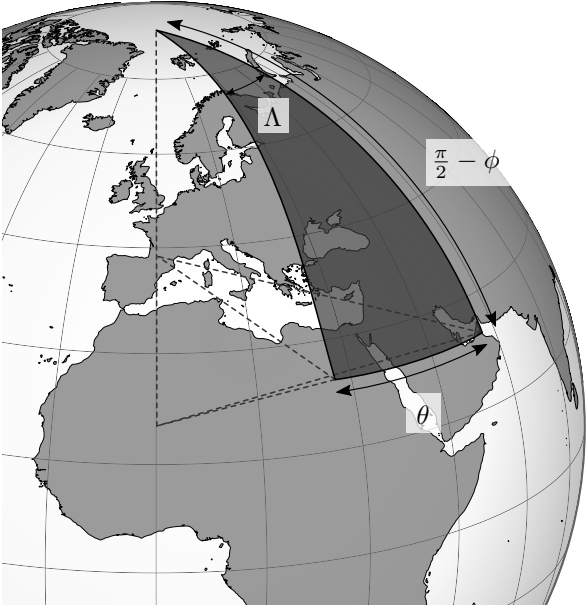
	\caption{Spherical triangle geometry for calculation of visible range in longitude $\Lambda$ from sensor half ground range angle $\theta$ at target latitude $\phi$.}
	\label{F:SphericalTrig}
	\end{minipage}
\end{figure}

From a given elevation angle constraint $\epsilon$, the half ground range angle $\theta$ can be calculated by considering the oblique triangle OSP.
\begin{equation}
\theta = \cos^{-1} \left( \frac{R_\phi}{r_s} \cos{\epsilon} \right) - \epsilon
\end{equation}
Alternatively, given the angular field of regard or half-cone boresight angle $\psi$ available to the sensor an intermediate angle $\gamma$ and the slant range $\rho$ to the edge of the sensor coverage area can first be calculated.
\begin{equation}
\sin{\gamma} = \sin{\left(\epsilon + \tfrac{\pi}{2}\right)} = \left( \frac{r_s \sin{\psi}}{R_{\phi}} \right)
\end{equation}
\begin{equation} \label{E:SlantRange}
\rho = R_{\phi} \cos{\gamma} + r_s \cos{\psi}
\end{equation}
The half ground range angle $\theta$ from the nadir can then be calculated using the sine-law.
\begin{equation} \label{E:FoR_Phi}
\sin{\theta} = \frac{\rho\sin{\psi}}{R_{\phi}}
\end{equation}
The surface dihedral angle $\tilde{\Lambda}$, analogous to the coverage in longitude of the sensor at the target latitude, is calculated from the ground range angle using the spherical trigonometry shown \cref{F:SphericalTrig}.
\begin{multline}
\cos\theta = \cos(\tfrac{\pi}{2}-\phi)\cos(\tfrac{\pi}{2}-\phi) \\
+ \sin(\tfrac{\pi}{2}-\phi)\sin(\tfrac{\pi}{2}-\phi)\cos{\Lambda}
\end{multline}
\begin{equation} \label{FoR_Eq}
\tilde{\Lambda}_\phi = 2\Lambda_\phi = 2\cos^{-1} \left(\frac{\cos{\theta} - \sin^2{\phi}}{\cos^2{\phi}}\right)
\end{equation}

\subsection{Calculation of Longitude of Successive Passes}
In order to find the maximum difference in time between any two passes over the same point, the crossing longitudes of the satellite(s) at the target latitude is required. A J2 perturbed orbit is assumed to include the rotation of the orbital plane due to the non-spherical potential of the Earth. First, the the nodal period $P_n$ of the orbit is calculated from the Keplerian period $P_k$ \cite{Vallado2013}.
\begin{equation} \label{E:KepPeriod}
P_k = 2\pi \sqrt{\dfrac{a^3}{\mu_E}}
\end{equation}
\begin{multline} \label{E:NodalPeriod}
P_n = P_k \bigg[ 1 + \frac{3 J_2 R_a}{4p} \bigg\{ \sqrt{1-e^2} \left(2 - 3\sin^2{i} \right) \\
+ \left(4 - 5\sin^2{i} \right) \bigg\} \bigg] ^{-1}
\end{multline}
Then drift in Right Ascension of Ascending Node (RAAN) due to J2 orbit regression is also calculated \cite{Vallado2013}. Extended expressions for the secular effect on RAAN by further zonal harmonics (J4, J6) can also be used. 
\begin{align} \label{E:dRAAN}
\delta\Omega = &-\frac{3}{2} n J_2 \left( \frac{R_a}{p} \right)^2 \cos{i} \nonumber \\ 
&+ \frac{3}{32} n J_2^2 \left( \frac{R_a}{p} \right)^4 \cos{i} \left[ 12-4e^2 - \left( 80+5e^2 \right) \sin^2{i} \right]
\end{align}
The drift in longitude of successive passes can subsequently be calculated by considering the rate of Earth rotation $\omega_E$, rate of nodal regression $\delta\Omega$, and nodal period $P_n$.
\begin{equation} \label{dLon}
\Delta\lambda = P_n \left( -\omega_E + \delta\Omega \right)
\end{equation}
The longitude $\lambda_{\phi}$ of the orbit ground track at the target latitude can be calculated by finding the value of the true anomaly $\nu$ at the target latitude using \cref{E:TrueAnom}. As the orbit will both ascend and descend over the target latitude in any given pass (for $i > \phi$), a pair of results will exist for both $\nu_\phi$ and $\lambda_{\phi}$ representing the location of the ascending and descending passes.
\begin{multline} \label{E:Longitude}
\lambda = \tan^{-1} \left\{ \frac{\cos{(\omega+\nu_\phi)}\sin{\Omega} + \sin{(\omega+\nu_\phi)}\cos{\Omega}\cos{i}}{\cos{(\omega+\nu_\phi)}\cos{\Omega} - \sin{(\omega+\nu_\phi)}\sin{\Omega}\cos{i}} \right\} \\
+ \frac{\nu_\phi}{2\pi}\Delta\lambda
\end{multline}
The longitude of successive passes at the target latitude up to a specified time can then be expressed as an arithmetic series in which the difference between consecutive terms is the drift in longitude $\Delta\lambda$.  
\begin{equation} \label{E:Pass_base}
\{\lambda_{j}\} = \lambda_{1} + \left(j-1\right)\Delta\lambda \quad \textrm{for} \ j = 1,2,3\dots \left\Vert \frac{T_{lim}}{P_n} \right\Vert
\end{equation}

\subsection{Constellations of Satellites}
Support for multi-satellite configurations is provided by calculating the offset in longitude between corresponding ascending or descending passes at the target latitude. The standard Walker constellation notation of $i:t/p/f$ is used, where $i$ is the inclination; $t$ is the total number of satellites; $p$ is the number of equally spaced planes; and $f$ is the relative spacing between satellites in adjacent planes. The number of satellite per plane $s$ can subsequently be defined by dividing the total number of satellites equally between the number of planes.
\begin{equation}
s=\frac{t}{p}
\end{equation}
The longitude of passes of satellites in multiple planes can be calculated by considering the angular separation in RAAN from a reference satellite and the relative spacing factor $f$ between satellites in each plane.
\begin{equation} \label{E:Pass_planes}
\{\lambda_{p}\} = \{\lambda_{j}\} + 2 \pi m \left( \frac{1}{p} + \frac{f}{t} \right) \quad \textrm{for} \ m = 1 ~ \textrm{to} ~ (p-1)
\end{equation}
Similarly, the longitude of corresponding passes of multiple satellites in a each plane can be calculated by considering the in-track spacing between the satellites and the rate of drift in longitude $\Delta\lambda$.
\begin{equation} \label{Pass_inplane}
\{{\lambda_{s}}\} = \{\lambda_{j}\} + \frac{l}{s}\Delta\lambda \quad \textrm{for} \ l = 1 ~ \textrm{to} ~ (s-1)
\end{equation}

Finally, the total set of passes by longitude in a given analysis period can be generated by combining the calculated pass sets.
\begin{equation}
\lambda_{\phi} = \lambda_j \cup \lambda_s \cup \lambda_p
\end{equation}

Alternatively, for non-symmetric constellations of satellites, each plane of the constellation can be considered individually with the possibility of variation in RAAN of each plane and in-plane spacing of the satellites.

\subsection{Determination of Visible Longitudes}

\begin{figure}[bt]
	\centering
	\def\svgwidth{110mm}
	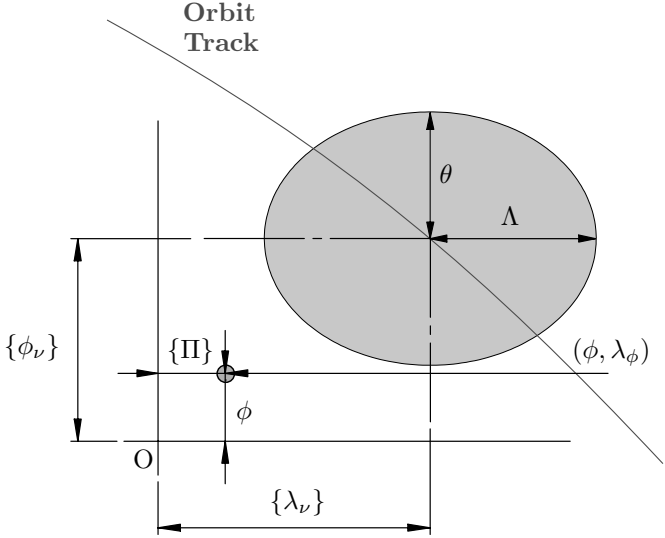
	\caption{Reference geometry used to determine whether a given point at the target latitude exists within the boundary of the representative sensor ground ellipse.}
	\label{F:Ellipse}
\end{figure}

A discretised grid of longitudes $\{\Pi\}$ about the target latitude is defined to assess the revisit performance of the given constellation configuration, orbit geometry, and sensor parameters.
First, the inequality expressed in \cref{E:InView} is used to indicate which of the longitudes in $\{\Pi\}$ at the target latitude are visible by a pass of longitude $\lambda_{\phi}$ due to the angular range of the sensor $\Lambda$.
\begin{equation} \label{E:InView}
\lambda_{\phi} - \{\Pi\} \leq \Lambda
\end{equation}
Second, the rotation of the Earth and the angle between the orbit track and the latitude of interest during each pass are considered using the projection of the sensor footprint on the Earth surface. On a equirectanglar projection of the Earth this footprint forms a complex distorted elliptical shape owing to the oblate spheroid shape of the Earth and convergence of longitudinal lines towards the poles. However, the geometry of a common ellipse can be assumed for simplicity up to high latitudes of interest ($<\ang{75}$). The inequality expression in \cref{E:Ellipse} indicates whether a point $(x,y)$ falls within the bounds of an ellipse of radii $(a,b)$ with origin at $(g,h)$.
\begin{equation} \label{E:Ellipse}
\frac{\left( x-g \right)^2}{a^2} + \frac{\left( y-h \right)^2}{b^2} \leq 1
\end{equation}
Due to the oblate spheroid shape of the Earth, the size and shape of this ellipse is dependant on the latitude of interest, orbital altitude, and sensor boresight angle. In the region of the target latitude the semi-major axis $a$ is equal to the half-angular range of the sensor $\Lambda$ and the semi-minor axis $b$ equal to the half ground range angle $\theta$. The coordinates of the origin of the ellipse $(\lambda_\nu,\phi_\nu)$ are defined by the latitudinal and longitudinal coordinates of the ground track of the satellite over a range above and below the target latitude. The corresponding range of longitudes $\lambda_{\nu}$ is calculated using \cref{E:TrueAnom,E:Longitude}. The inequality to determine if any given longitude is visible during a given pass of the satellite, illustrated in \cref{F:Ellipse}, can therefore be expressed by \cref{E:ViewEllipse}.
\begin{equation} \label{E:ViewEllipse}
\frac{\left( \{\Pi\}-\lambda_{\nu} \right)^2}{\Lambda^2} + \frac{\left( \phi-\{\phi_{\nu}\} \right)^2}{\theta^2} \leq 1
\end{equation}
%

By evaluation of \cref{E:InView,E:ViewEllipse} for the longitude of each pass $\lambda_{\phi,j}$ at each latitude of interest, a list of accesses for each longitude on the discretised grid $\{\Pi\}$ is established. Accuracy in the calculation of revisit time is maintained by using the true anomaly to calculate the start and end time of each access. The maximum gap in time between two consecutive accesses for any longitude in $\{\Pi\}$ represents the maximum revisit time for the given orbit and sensor configuration.

\section{Results}
The accuracy of the presented method is first assessed by comparison to numerical simulations generated by STK and previously published results in the literature. Results for single and multi-satellite MRT are then presented for ranges of orbital altitude, inclination, and sensor FoR.

In order to limit the required memory for computation of revisit time using the presented method, a limit on the resolution in longitude $\{\Pi\}$ and discretisation of each pass over the target latitude $\{\phi_\nu\}$ can be implemented. Similarly, the number of passes can be constrained, limiting the length of the analysis period. For the following cases a resolution of \ang{0.1} in longitude is used and each pass over the equator is considered using a set of \num{1000} points.

For representative comparison of the computational efficiency, the developed implementation in MATLAB has an average run-time per function call in serial computation mode of less than \SI{0.80}{\second} on a Intel Core i7-4770 3.40GHz workstation. Parallel computing methods can be used to improve this performance significantly by increasing the number of computations which can be performed in a given period of time.

\subsection{Validation}
Validation of the presented method is first performed by comparing single satellite revisit metrics to those obtained using the orbital simulation software STK. In each case, STK was set-up and run using the programming interface (no GUI) to reduce user input. The ``J2Perturbation'' analytical propagation method was used to match the implemented orbit perturbations of the method and an analysis period of 60 days set, sufficient to identify if a case demonstrates singularity due to a near repeat ground track pattern.

 A $2^3$ fractional factorial analysis of cases of equatorial revisit time in LEO covering variation in orbital altitude, inclination and sensor angle is performed. A pair of additional cases for representative SSO orbits are also included, yielding a total of \num{10} cases. \cref{T:Validation1} shows the results of the factorial analysis and associated error for each case. The greatest difference in MRT is \SI{0.01}{\hour}, corresponding to a absolute percentage error of \SI{0.17}{\%}. An absolute error of less than one minute with the numerical simulation is demonstrated in all cases.

\begin{table*}[!tpb]
	\newcolumntype{R}{>{\raggedleft\arraybackslash\hspace{0pt}}m{0.075\linewidth}}
	\newcolumntype{N}{>{\raggedleft\arraybackslash\hspace{0pt}}m{0.1\linewidth}}
		\caption{Validation of MRT calculation with varying altitude $h$, inclination $i$, and elevation angle constraint $\epsilon$.} \label{T:Validation1}
		\centering
		\begin{tabular}{RRRNNNN}
		\toprule
		\multirow{2}{*}[-2pt]{\textbf{\boldmath$h$ [km]}} & \multirow{2}{*}[-2pt]{\textbf{\boldmath$i$ [deg]}} & \multirow{2}{*}[-2pt]{\textbf{\boldmath$\epsilon$ [deg]}} & \multicolumn{3}{c}{\bf MRT [hours]} & \multirow{2}{*}[-2pt]{\textbf{\% Error}}\\
		\cmidrule{4-6}
		&  &  & {\bf Method} & {\bf STK} & {\bf Error} & \\
		\midrule
		400 & 20 & 10 & 9.78  & 9.78  & -0.00 & 0.03 \\
		400 & 20 & 40 & 24.65 & 24.65 & -0.00 & 0.01 \\ 
		400 & 60 & 10 & 13.08 & 13.08 & -0.00 & 0.00 \\
		400 & 60 & 40 & 59.37 & 59.37 & -0.00 & 0.00 \\
		800 & 20 & 10 & 5.32  & 5.32  & -0.01 & 0.17 \\
		800 & 20 & 40 & 10.79 & 10.79 & -0.00 & 0.01 \\ 
		800 & 60 & 10 & 10.76 & 10.76 &  0.00 & 0.02 \\
		800 & 60 & 40 & 23.48 & 23.48 & -0.00 & 0.00 \\
		550 & 97.59 & 20 & 109.30 & 109.30 & -0.00 & 0.00 \\
		700 & 98.19 & 30 & 35.38 & 35.38 & -0.00 & 0.00 \\  
		\bottomrule
		\end{tabular}
		\end{table*}

The second validation process seeks to investigate the behaviour of the developed method with varying latitude of interest. This variable introduces variation in the sensor footprint characteristics due to the oblate spheroid nature of the Earth. For these cases, a SSO orbit of altitude \SI{500}{\kilo\meter} and inclination of \ang{97.41} is used and a elevation angle constraint of \ang{30} specified. The target latitude is varied from the equator to \ang{75} in increments of \ang{5}. The results with errors are shown in \cref{T:Validation2}. The maximum error of \SI{0.01}{\hour} or \SI{0.05}{\%} is shown to occur at \ang{75} and \ang{80}, the highest latitudes of interest. This is attributable to the assumption of an elliptical sensor footprint and the departure from this shape which occurs at high latitudes.

\begin{table}[!tbp]
	\caption{Validation of MRT calculation with varying latitude of interest for SSO at \SI{500}{\kilo\meter} altitude ($i =$ \ang{97}, $\epsilon =$ \ang{30}).} \label{T:Validation2}
	\centering
	\begin{tabular}{crrrr}
	\toprule
	\textbf{Latitude} & \multicolumn{3}{c}{\bf MRT [hours]} & \multirow{2}{*}[-2pt]{\textbf{\% Error}} \\
	\cmidrule{2-4}
	\textbf{[deg]}&  {\bf Method} & {\bf STK} & {\bf Error} & \\
	\midrule
	0  &	72.59	&	72.59	&	0.00	&	0.00	\\
	5  &	84.38	&	84.38	&	0.00	&	0.00	\\ 
	10 &	60.66	&	60.65	&	0.00	&	0.00	\\
	15 &	60.60	&	60.60	&	0.00	&	0.00	\\ 
	20 &	36.88	&	36.88	&	0.00	&	0.00	\\
	25 &	36.83	&	36.83	&	0.00	&	0.00	\\ 
	30 &	23.65	&	23.65	&	0.00	&	0.00	\\ 
	35 &	35.78	&	35.78	&	0.00	&	0.00	\\ 
	40 &	35.83	&	35.83	&	0.00	&	0.00	\\ 
	45 &	35.88	&	35.88	&	0.00	&	0.00	\\ 
	50 &	25.23	&	25.23	&	0.00	&	0.01	\\ 
	55 &	14.46	&	14.46	&	0.00	&	0.02	\\
	60 &	14.41	&	14.41	&	0.00	&	0.00	\\
	65 &	14.36	&	14.36	&	0.00	&	0.00	\\
	70 &	14.32	&	14.32	&	0.00	&	0.02	\\
	75 &	14.28	&	14.28	&	0.01	&	0.05	\\
	80 & 	14.24	& 	14.25	& 	0.01	& 	0.05	\\
	\bottomrule
	\end{tabular}
	\end{table}


\subsection{Comparison to Published Data}
MRT calculations for constellations of satellites can be compared to results in the literature presented by \citet{Lang2003} and \citet{Ulybyshev2014}. Revisit time is calculated at the equator for a basic Walker constellation configuration with three satellites. Three cases with varying altitude, inclination, elevation angle constraint are presented. The result obtained using numerical simulation by STK for these cases is also shown. The results shown in \cref{T:Validation3} demonstrate a significant increase in accuracy of the presented method in comparison to other existing analytical processes.

\begin{table}[!tbp]
		\caption{Validation of MRT calculation for constellations of different inclination and elevation angle constraint.} \label{T:Validation3}
		\centering
		\begin{tabular}{lrrr}
		\toprule
		{\bf Parameter} \\
		\midrule
		Inclination, $i$ & \ang{90} & \ang{86} & \ang{96} \\
		Configuration, $t/p/f$ & $3/3/0$ & $3/3/0$ & $3/3/1$ \\
		Altitude [km] & 700 & 1100 & 1500 \\
		Elevation, $\epsilon$ & \ang{0} & \ang{10} & \ang{20} \\
		\midrule
		{\bf Solution} & \multicolumn{3}{c}{\bf MRT [hours]} \\
		\midrule
		\citet{Lang2003} & 2.60 & 4.35 & 3.78\\
		\citet{Ulybyshev2014} & - & 4.46 & -\\
		STK & 2.30 & 4.25 & 3.38\\
		Method & 2.30 & 4.25 & 3.38\\
		\bottomrule
		\end{tabular}
		\end{table}

\subsection{Maps of Single Satellite and Constellation Revisit Time}
\begin{figure}[htb] 
	\centering
	\includegraphics[width=\linewidth]{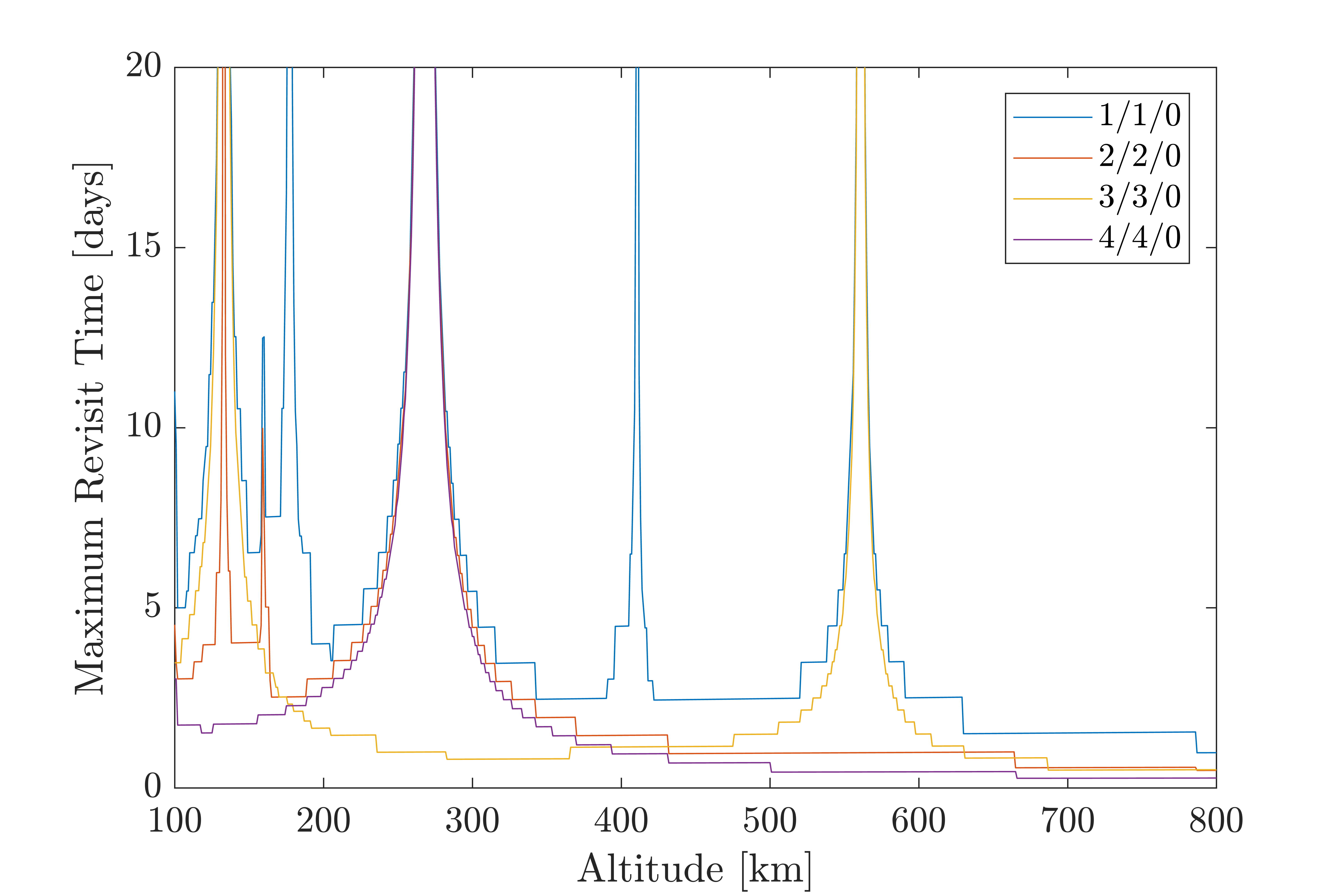}
	\caption{MRT computation for varying Sun-synchronous orbit altitude and constellation configuration for a target latitude of \ang{40} and sensor boresight half-cone angle of \ang{45}.}
	\label{F:SSO_Constellation}
	\end{figure}

\begin{figure}[htb] 
	\centering
	\includegraphics[width=\linewidth]{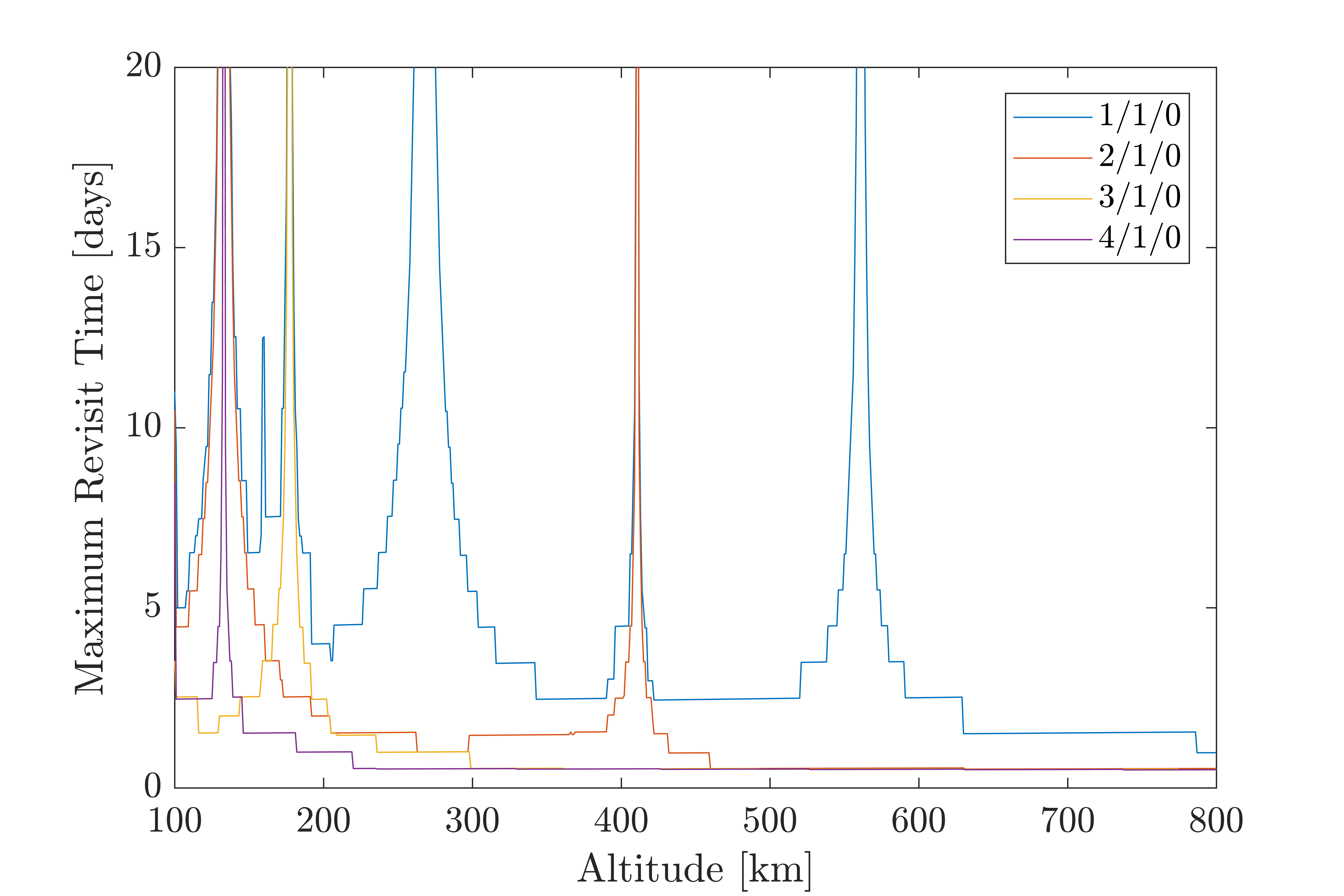}
	\caption{MRT computation for varying Sun-synchronous orbit altitude and number of satellites equispaced in a single orbital plane for a target latitude of \ang{40} and sensor boresight half-cone angle of \ang{45}.}
	\label{F:SSO_SingleOrbit}
	\end{figure}

MRT maps are provided for a single satellite in a single orbital plane and for some constellations configurations characterised by various number of equally spaced planes. During the first stage of the implementation, a sun-synchronous aspect of the orbit was assumed and the values of the target latitude and the FoR angle were set equals to \ang{40} and \ang{45} respectively. \cref{F:SSO_Constellation} shows how temporal resolution introduces some constraints on the altitude windows in which LEO satellites can be operated. The low MRT achievable for the range 600-800 km makes these altitude windows suitable for most Earth Observation missions. Generally speaking, for certain ranges, small variations in altitude can eventually result in non-negligible differences in temporal resolution performance. However, it is interesting to notice how Sun-synchronous satellite constellations consisting of an odd number of planes, any one of which occupied by a single satellite, provide significant improvement for certain lower altitudes windows, granting comparable performance in terms of temporal resolution to higher altitudes. As expected, augmenting the number of operative satellites in a single orbital plane, results in improved temporal resolution with better results obtained with increasing number of satellites employed, demonstrated in \cref{F:SSO_SingleOrbit}.

\begin{figure}[htb] 
	\centering
	\includegraphics[width=\linewidth]{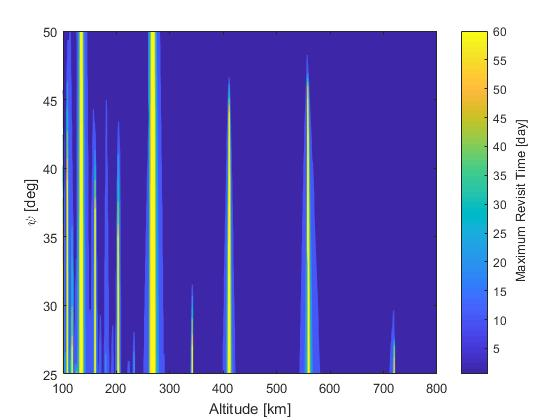}
	\caption{Contour plot of MRT for varying Sun-synchronous orbit altitude and sensor boresight half-cone angle for a single satellite and target latitude of \ang{40}.}
	\label{F:Contour_SSO_110}
	\end{figure}

\begin{figure}[htb] 
	\centering
	\includegraphics[width=\linewidth]{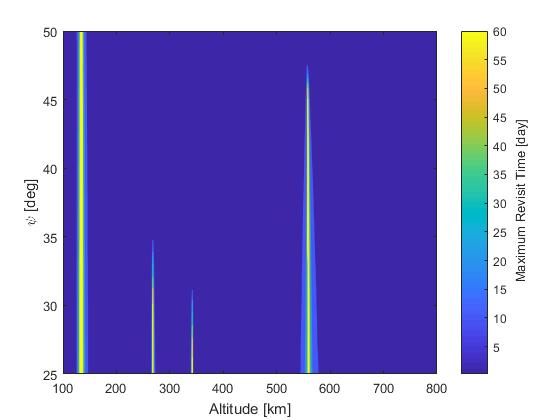}
	\caption{Contour plot of MRT for varying Sun-synchronous orbit altitude and sensor boresight half-cone angle for a 3/3/0 constellation and target latitude of \ang{40}.}
	\label{F:Contour_SSO_330}
	\end{figure}

Results obtained for a single Sun-synchronous orbiting satellite according to varying boresight half-cone angle values and altitudes, shown in \cref{F:Contour_SSO_110} confirm, as expected, that better results are achievable when higher FoR angles are employed. At higher altitudes, the impact on temporal resolution should be more relevant because the ground area sensed increases for the same sensor properties. For some combinations of low altitudes and narrow half-cone boresight angle, shown in \cref{F:Contour_SSO_110}, the analysis period was exceeded and thus \SI{100}{\%} global coverage was not achieved. Similar results for a 3/3/0 Walker constellation, shown in \cref{F:Contour_SSO_330}, demonstrate improved revisit metrics for LEO operations employing narrow FoR at certain altitude windows. Low but not necessarily continuous revisit time is also achievable for a wide range of altitudes when high sensor boresight half-cone angle is selected.

Major contributions to temporal resolution can also be guaranteed by off-nadir pointing capability when direct coverage between adjacent ground tracks is not achieved. The method does not directly address the effect of off-nadir pointing angle on MRT computation, but comparable results can be obtained by increasing the FoR used.

\begin{figure}[htb] 
	\centering
	\includegraphics[width=\linewidth]{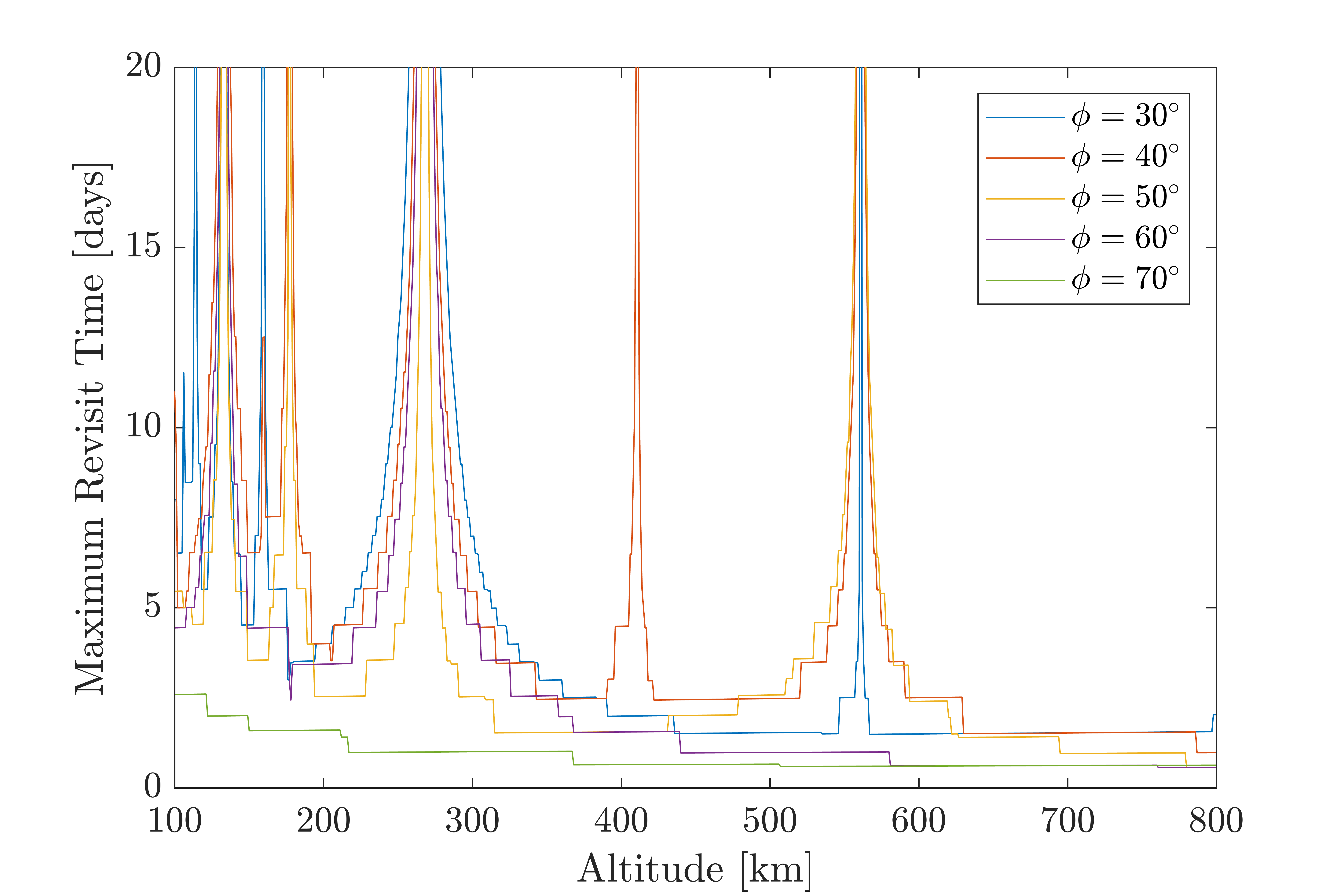}
	\caption{MRT computation for varying Sun-synchronous orbit altitude and target latitude for a single satellite with sensor boresight half-cone angle of \ang{45}.}
	\label{F:SSO_110_Latitude}
	\end{figure}

For non-Earth-synchronous or repeat ground track orbits, the projected ground track of a satellite appears to move with respect to the Earth between subsequent orbits. As a consequence, consecutive passages over the equator are separated in longitude, dependent on both the orbital period and the rotation rate of the Earth \cite{Wertz1999}. For target latitudes approaching the poles (eg $\phi=\ang{70}$), the distance between the lines of longitude diminishes and consequently the gaps which separate adjacent ground tracks from each other at the target latitude are reduced for the same sensor properties, resulting in the expected improvement in MRT demonstrated in \cref{F:SSO_110_Latitude}.

\begin{figure}[htb]
	\centering
	\includegraphics[width=\linewidth]{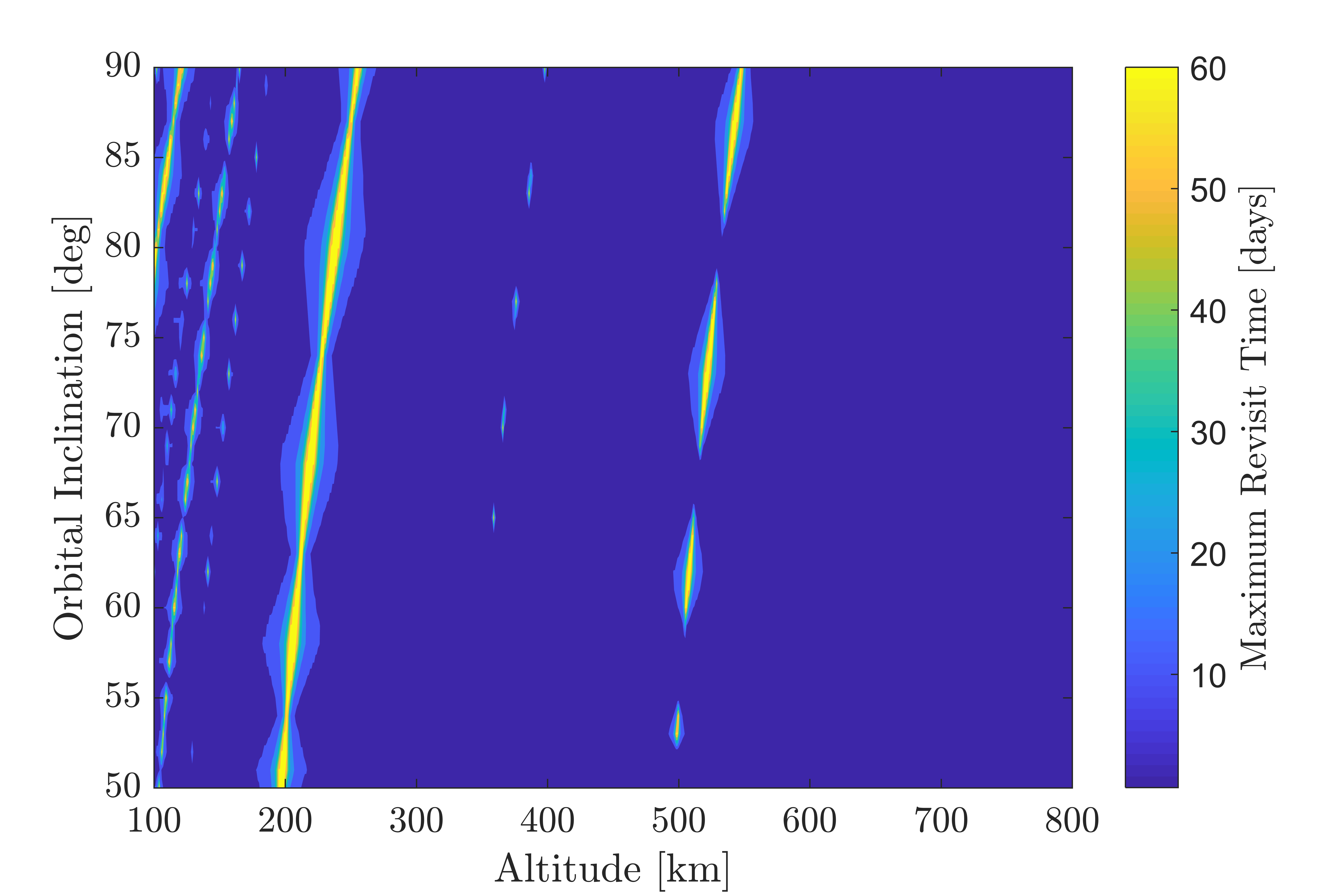}
	\caption{Contour plot of MRT for varying non-SSO orbit altitude and inclination for a single satellite with sensor boresight half-cone angle of \ang{45} and target latitude of \ang{40}.}
	\label{F:Contour_Inc_110_FoR_45}
	\end{figure}

\begin{figure}[htb]
	\centering
	\includegraphics[width=\linewidth]{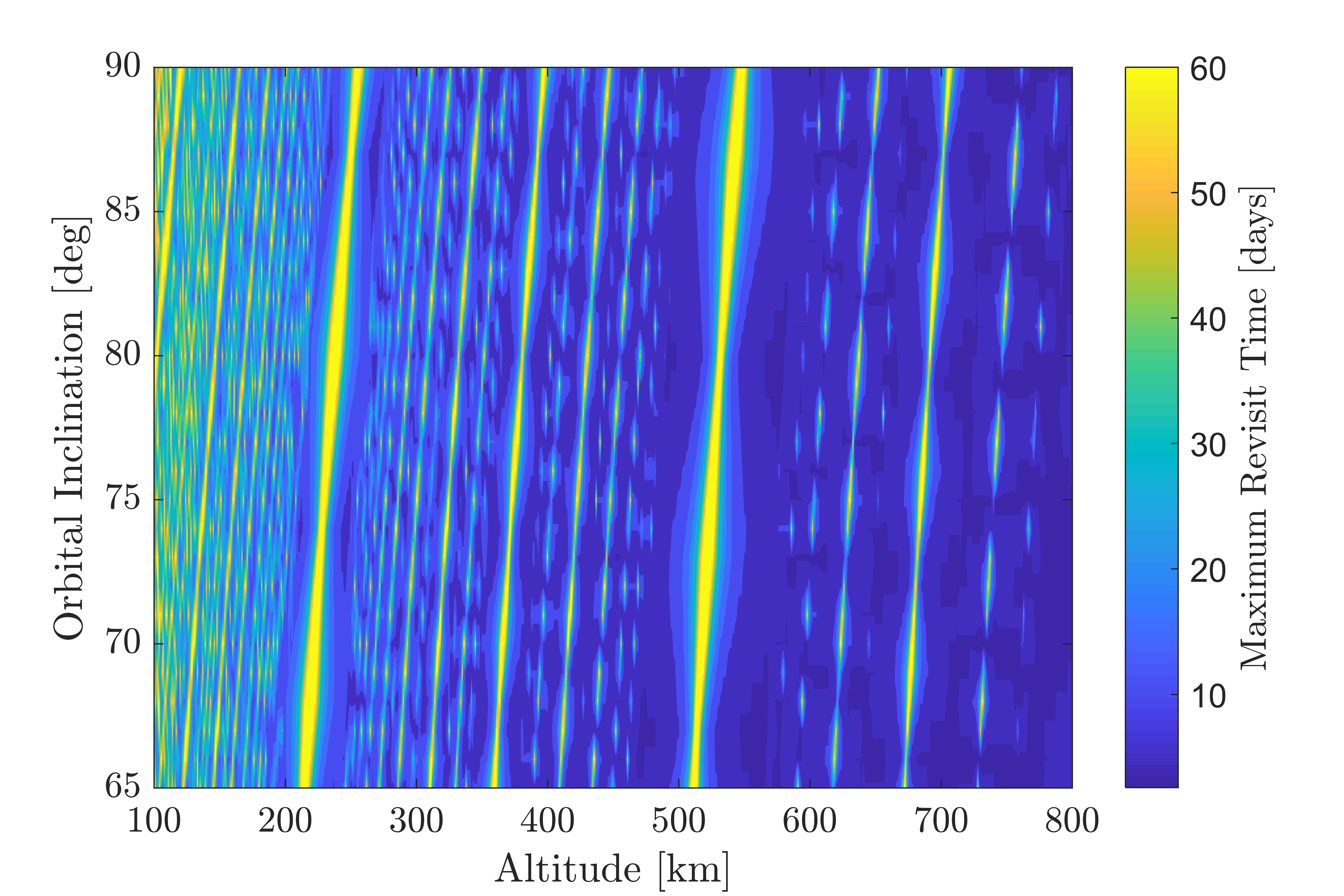}
	\caption{Contour plot of MRT for varying non-SSO orbit altitude and inclination for a single satellite with sensor boresight half-cone angle of \ang{15} and target latitude of \ang{40}.}
	\label{F:Contour_Inc_110_FoR_15}
	\end{figure}

The impact of variation in angular FoR and orbit inclination for non-SSOs on temporal resolution is interesting. For a given altitude, the MRT is not constant for different inclinations due to the variation in ground-track shift, demonstrated in \cref{F:Contour_Inc_110_FoR_45} and \cref{F:Contour_Inc_110_FoR_15}. These results also demonstrate that restriction of the angular FoR, by either sensor selection or pointing capability, necessitates careful selection of the orbit inclination in coordination with the altitude to obtain optimal revisit metrics even at altitudes greater than \SI{600}{\kilo\meter}. However, for low FoR certain altitude windows may need to be avoided, eg less than \SI{200}{\kilo\meter} and approximately \SIlist{350;500}{\kilo\meter}, where no inclination can be found that guarantees low revisit. Furthermore, for EO satellites which will experience orbital decay during their operational lifetime, consideration of variation in inclination may also be required in order to maintain the optimal MRT.

\section{Conclusions}
This paper presents a semi-analytical method developed for calculating revisit time for single satellite and constellations in circular orbit with discontinuous coverage. For the selected target latitude, the method provides a database where revisit metrics are calculated for varying orbit altitudes, viewing or elevation angles, and inclinations (in case of non-SSO) establishing a correlation between the longitude of the projected passages and the instantaneous coverage of the sensor. Preliminary orbit and sensor geometry computation is carried out to determine the sensor coverage in longitude at the target latitude. The maximum time difference between any two consecutive passes over a selected target is then determined through the computation of the longitudes crossing the latitude of interest according to the satellite's motion. Orbital perturbations associated with the Earth's oblateness are considered through the J2 zonal spherical harmonics. Determination of visible longitudes is finally performed for varying orbit configuration, sensor properties and orbit geometry, defining a discretised grid of longitudes in the proximity of the selected latitude. The longitudes visible at each pass are those which fall within the sensor footprint projection on the Earth's surface. For a matter of simplicity, the footprint is approximated with an ellipse, which is demonstrated to provide accurate results for the latitude range of interest ($<\ang{75}$).

The method accuracy was validated by comparison to results published in literature and STK numerical simulations. A maximum error of less than a minute was registered in all cases, demonstrating the accuracy of the presented method and the meaningful improvements achieved in revisit metrics computation compared to other analytical routines. The computational efficiency of the method was also shown to be significantly better than widely available commercial (numerical-based) software and similar to that of other previously published information on analytical methods, albeit using more modern hardware. The combination of these performance improvements will support better early design phase analysis and more complete design optimisation processes.

The capability of the method was demonstrated through the generation of plots that yield information on MRT characteristics for varying target latitudes, sensing capabilities, orbital parameters and configurations for both single satellite and Walker constellation architectures. These plots on their own and the supporting data can provide valuable reference for Earth observation system designers in the early phases of a mission development lifecycle.

Future developments of this method may include extensions to single satellite and constellations in elliptical orbits as well as non-symmetrical satellite constellations. Off-nadir pointing capabilities could also be directly addressed, and the related impact on MRT computation discussed. Calculation of revisit time metrics considering time of night/day or lighting conditions would also enable particular application for optical Earth observation sensors. Finally, more complex sensor shapes could be investigated and different projection geometries could be studied to ensure accuracy of the method at higher target latitudes.

\section*{Acknowledgements}
This work was supported by the Doctoral Training Partnership (DTP) between The University of Manchester and the UK Engineering and Physical Sciences Research Council (EPSRC) under grant EP/M506436/1.

\bibliography{2016_Revisit}

\end{document}